\documentclass[epsfig]{ws-procs9x6}
\begin{document}

\title{$Q^2$-evolution of $\Delta N \gamma$ form factors
up to $4~(GeV/c)^2$ from JLab data }

\author{I.G.Aznauryan}

\address{Yerevan Physics Institute, \\
Alikhanian Brothers St. 2, \\ 
Yerevan, 375036 Armenia \\ 
E-mail: aznaury@jlab.org}

\maketitle

\abstracts{
We present the results on the ratios
$E_{1+}^{(3/2)}/M_{1+}^{(3/2)}$ and
$S_{1+}^{(3/2)}/M_{1+}^{(3/2)}$
for the $\gamma^* N \rightarrow \Delta(1232)$ transition
at $Q^2\leq 4~(GeV/c)^2$
extracted from the  $p(e,e'p)\pi^0$ cross section using two
approaches: dispersion relations and  modified version
of unitary isobar model. The obtained results are in good agreement
with the results of other analyses
obtained using truncated
multipole expansion at
$Q^2=0.4,~0.525,~0.65,~0.75,~0.9,~1.15,~1.45,~1.8~(GeV/c)^2 $ 
and within dynamical and unitary isobar models at
 $Q^2=2.8,~4~(GeV/c)^2 $. According to obtained results
the ratio $E_{1+}^{(3/2)}/M_{1+}^{(3/2)}$
remains small in all investigated region of $Q^2$
with very unclear tendency  to cross zero above
$2~(GeV/c)^2$. The absolute value of the
 ratio $S_{1+}^{(3/2)}/M_{1+}^{(3/2)}$
 is clearly increasing with increasing $Q^2$,
while it should be a constant value in the pQCD asymptotics.
So, at $Q^2\leq4~(GeV/c)^2$   
there is no evidence of approaching
pQCD regime for these ratios.
None of the soft approaches gives satisfactory
description of the obtained results.
}

\section{Introduction}
It is known that for about 20 years the question: which is the scale
of transition from soft to hard mechanism of QCD
in exclusive processes,
is the subject of controversy. Detail discussion
of this problem can be found, for example, in  
papers \cite{1,2,3}. The point of view,
that this scale should be large, i.e. much larger than now
available $Q^2$, is based mainly on the utilization of asymptotic
wave function  and is confirmed by the results obtained
using local quark-hadron duality  \cite{4}. On the other
hand there are arguments, based mainly
on the utilization of the Chernyak-Zhitnitsky wave function
\cite{5},  that hard mechanism of QCD
can be observed at quite small $Q^2$. Experimental data on proton elastic 
form factors and form factors for the second
and third resonance peaks extracted from
inclusive data  \cite{3}, indeed,
manifest the features which are characteristic of pQCD
starting with very small $Q^2$, about $2-3~GeV^2$.
However,  for
$\gamma^* N \rightarrow \Delta(1232)$ the strong
numerical suppression of the leading order amplitude
is obtained using the wave functions of CZ type
\cite{6}.
By this reason, in distinction to other form factors,
the hard mechanism is expected
for $\gamma^* N \rightarrow \Delta(1232)$
at much higher $Q^2$.
Information on the $Q^2$- evolution
of the ratios
$E_{1+}^{(3/2)}/M_{1+}^{(3/2)}$,  
$S_{1+}^{(3/2)}/M_{1+}^{(3/2)}$
for the  $\gamma^* N \rightarrow \Delta(1232)$
transition
will allow to check this expectation, because
 the transition from soft to hard mechanism
is characterized by a striking change
in the behaviour of
$E_{1+}^{3/2}/M_{1+}^{3/2}$,
$S_{1+}^{3/2}/M_{1+}^{3/2}$ from
\begin{equation}
E_{1+}^{3/2}/M_{1+}^{3/2}\cong 0, ~
S_{1+}^{3/2}/M_{1+}^{3/2}\cong 0
\label{eq:1}
\end{equation}
at $Q^2=0$ to
\begin{equation}
E_{1+}^{3/2}/M_{1+}^{3/2}=1, ~
S_{1+}^{3/2}/M_{1+}^{3/2}=const
\label{eq:2}
\end{equation}
in the pQCD regime.
By this reason, investigation of the $Q^2$-evolution of the ratios
$E_{1+}^{(3/2)}/M_{1+}^{(3/2)}$,
$S_{1+}^{(3/2)}/M_{1+}^{(3/2)}$ 
is very informative for understanding of the mechanisms
and the scale of transition to pQCD regime.

In this report, we present the results on the ratios
$E_{1+}^{(3/2)}/M_{1+}^{(3/2)}$,
$S_{1+}^{(3/2)}/M_{1+}^{(3/2)}$
for the  $\gamma^* N \rightarrow \Delta(1232)$
transition at $Q^2=0.4,~0.525,~0.65,~0.75,~0.9,~1.15,~1.45,~1.8,~2.8,~4~(GeV/c)^2 $.
These results are
extracted from the JLab data  \cite{7,8}
on $p(e,e'p)\pi^0$ cross section using two approaches:
dispersion relations and modified version
of the unitary isobar model of Ref. \cite{9}. The detail description
of the approaches is done in  Ref. \cite{10}.

\section{Results and discission}

\begin{figure}[th]
\centerline{\epsfxsize=11cm\epsfysize=11cm\epsfbox{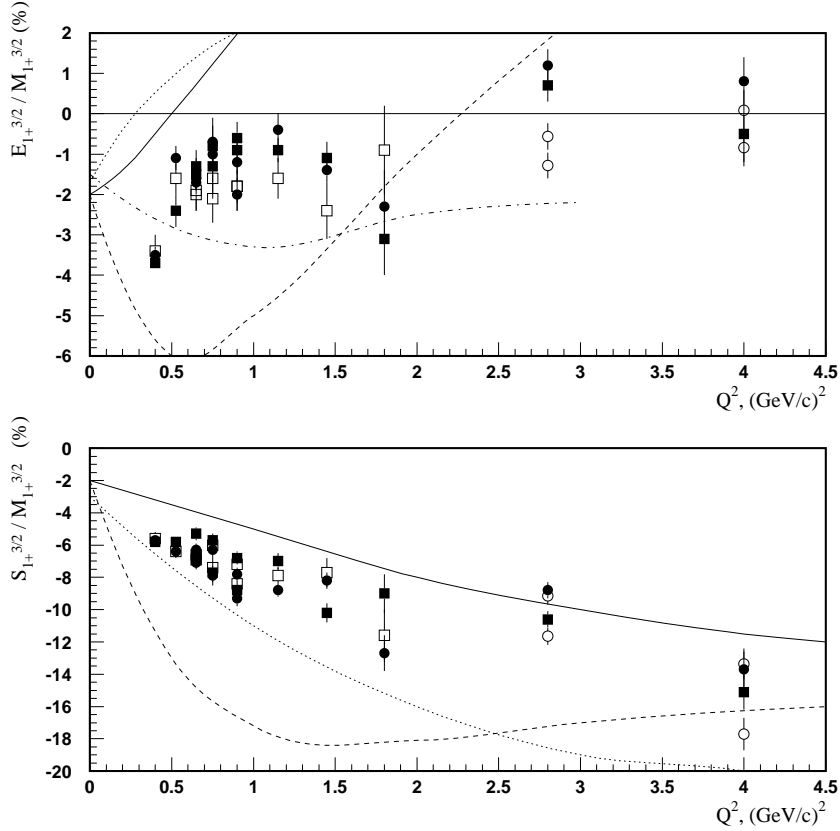}}
\caption{
The  ratios
$E_{1+}^{(3/2)}/M_{1+}^{(3/2)}$,
$S_{1+}^{(3/2)}/M_{1+}^{(3/2)}$
for the  $\gamma^* N \rightarrow \Delta(1232)$
transition extracted from the  $p(e,e'p)\pi^0$ 
cross section ${}^{7,8}$.
Results of our analysis are obtained using dispersion
relations (full circles) and our modified version
of unitary isobar model (full squares). Open squares
correspond to truncated multipole analysis  ${}^{8}$,
open  circles are obtained within dynamical and unitary isobar models
in Ref. ${}^{11}$. The predictions
of the light-cone relativistic quark
models are presented by solid ${}^{12}$ and dotted ${}^{13}$
curves, dashed-dotted curve corresponds to relativized
quark model ${}^{14}$, and dashed curves are
the results of Ref. ${}^{15}$.
\label{inter}}
\end{figure}

The obtained results for the ratios 
$E_{1+}^{(3/2)}/M_{1+}^{(3/2)}$,
$S_{1+}^{(3/2)}/M_{1+}^{(3/2)}$
are presented in  Figure~\ref{inter}. 

In this Figure the results obtained by truncated
multipole analysis of at 
$Q^2=0.4,~0.525,~0.65,~0.75,~0.9,~1.15,~1.45,~1.8~(GeV/c)^2 $ \cite{9}
and using dynamical and unitary isobar models at
 $Q^2=2.8,~4~(GeV/c)^2 $ \cite{11} are also presented.

Let us note, that presented results correspond
to the so called "dressed" $\gamma N \Delta$ vertex,
i.e. are extracted from the whole magnitudes of multipoles
$M_{1+}^{(3/2)}$, $E_{1+}^{(3/2)}$, $S_{1+}^{(3/2)}$
at the resonance position, where $\delta_{1+}^{(3/2)}=\pi/2$.
Extraction of "bare" multipoles can be made
only using models and is model-dependent.

Note also, that there are two sets of results at
$Q^2=0.65,~0.75,~0.9~(GeV/c)^2 $, which is connected
with two kinds of measurements of 
$p(e,e'p)\pi^0$ cross section in \cite{8}
at different energies of initial electron.

From  Figure~\ref{inter} it is seen that the results
obtained using different approaches agree with each other.
The ratio $E_{1+}^{(3/2)}/M_{1+}^{(3/2)}$
remains small in all $Q^2$ region up to $4~(GeV/c)^2$,
revealing some tendency to cross zero above
$Q^2>2~(GeV/c)^2$. However, this tendency
is not clear, and only measurements at higher
$Q^2$ can clear up the situation here.
The absolute value of the 
 ratio $S_{1+}^{(3/2)}/M_{1+}^{(3/2)}$
 is clearly increasing with increasing $Q^2$,
and does not reveal the tendency of approaching the
constant value. So, at $Q^2\leq4~(GeV/c)^2$
there is no indication of approaching
pQCD regime for the ratios
 $E_{1+}^{(3/2)}/M_{1+}^{(3/2)}$
and  $S_{1+}^{(3/2)}/M_{1+}^{(3/2)}$.

Therefore, it is reasonable to conclude, that the behaviour
of the ratios 
 $E_{1+}^{(3/2)}/M_{1+}^{(3/2)}$
and  $S_{1+}^{(3/2)}/M_{1+}^{(3/2)}$ at  $Q^2\leq4~(GeV/c)^2$
is related to the soft mechanisms.
By this reason in  Figure~\ref{inter} 
the predictions based on the soft approaches
are presented. These are light-cone relativistic quark
model predictions \cite{12,13}, predictions of relativized
quark model \cite{14}, and the results of Ref. \cite{15}
obtained by interpolation between very low and very high
$Q^2$. Both light-cone 
 relativistic quark
models give qualitatively  good description of the ratio
 $S_{1+}^{(3/2)}/M_{1+}^{(3/2)}$,
but in the case of  $E_{1+}^{(3/2)}/M_{1+}^{(3/2)}$
they contradict the obtained results.
In contrast with this,  the relativized
quark model predictions \cite{14}
for $E_{1+}^{(3/2)}/M_{1+}^{(3/2)}$
are close to the results extracted from experimental data,
however, in the case of $S_{1+}^{(3/2)}/M_{1+}^{(3/2)}$
the prediction of this model:
$S_{1+}^{(3/2)}/M_{1+}^{(3/2)}\simeq 0$,
 disagrees with these results. 
The predictions of Ref. \cite{15} contradict
the obtained results for both ratios. So,
none of the soft approaches
gives satisfactory description of the ratios
 $E_{1+}^{(3/2)}/M_{1+}^{(3/2)}$
and  $S_{1+}^{(3/2)}/M_{1+}^{(3/2)}$ in the 
investigated region of $Q^2$.

\section{Conclusion}
In summary, we have analysed JLab data on  $p(e,e'p)\pi^0$ cross section
at $Q^2\leq 4~(GeV/c)^2$
in the $\Delta(1232)$ resonance region within two approaches:
dispersion relations and modified version of unitary
isobar model. As a result, we have obtained information on $Q^2$-evolution
of the ratios  $E_{1+}^{(3/2)}/M_{1+}^{(3/2)}$
and  $S_{1+}^{(3/2)}/M_{1+}^{(3/2)}$, which are of
interest for understanding the scale and mechanisms
of transition from soft to hard regime of QCD
in exclusive processes.
The obtained results show that there is no evidence yet
of approaching pQCD regime for 
the $\gamma^* N \rightarrow \Delta(1232)$ transition,
and for investigation of the scale, where hard mechanism
for this transition begin to work, measurements
at higher $Q^2$ are needed.
We have compared the results on  $E_{1+}^{(3/2)}/M_{1+}^{(3/2)}$
and  $S_{1+}^{(3/2)}/M_{1+}^{(3/2)}$ with existing predictions
obtained within soft approaches. It turned out,
that none of the soft approaches gives simultaneously
satisfactory description of these ratios.
This means, that more detail investigations of soft mechamisms
for  $\gamma^* N \rightarrow \Delta(1232)$ are necessary.

\section{Acknowledgements} 
I am thankful to V. D. Burkert, C. E. Carlson,  A. V. Radyushkin
and  V. L. Chernyak for useful discussions. I express my gratitude
for the hospitality at Jefferson Lab where this work
was done.

\end{document}